\documentclass[journal]{IEEEtran}
\usepackage{cite}
\usepackage{amsmath,amssymb,amsfonts}
\usepackage{graphicx}
\usepackage{textcomp}
\usepackage{xcolor}

\usepackage{hyperref}
\usepackage{amsmath}
\usepackage{autobreak}
\usepackage{amsfonts}
\usepackage{bm}
\usepackage{bbm}
\usepackage{algorithm}  
\usepackage{algorithmicx}  
\usepackage{algpseudocode}
\usepackage{stfloats}
\usepackage{cuted}
\usepackage{color}
\usepackage{subfigure}
\usepackage{cite}
\usepackage{booktabs} % 导入三线表需要的宏包
\usepackage{soul}

\definecolor{red}{RGB}{0,0,0}
\newcommand{\red}[1]{{\color{red}#1}}
\newcommand{\diag}{\text{diag}}

\makeatletter
\renewcommand{\maketag@@@}[1]{\hbox{\m@th\normalsize\normalfont#1}}%
\makeatother

\newcommand{\setParDis}{\setlength {\parskip} {0.1cm} }
\ifCLASSINFOpdf

\else

\fi

\begin{document}
	
	\title{Hybrid NOMA assisted Integrated Sensing and Communication via RIS}

	\author{Wanting Lyu,~Yue Xiu,~Xinyang Li,~Songjie Yang,~Phee Lep Yeoh,~\IEEEmembership{Member,~IEEE}, \\~Yonghui Li,~\IEEEmembership{Fellow,~IEEE},~Zhongpei Zhang,~\IEEEmembership{Member,~IEEE} 
		\vspace*{-0.8cm}
		\thanks{
			Wanting Lyu, Yue Xiu, Xinyang Li, Songjie Yang and Zhongpei Zhang are with National Key Laboratory of Science and Technology on Communications, University of Electronic Science and Technology of China, Chengdu 611731, China (E-mail: lyuwanting@yeah.net; xiuyue12345678@163.com;  lixinyang829@gmail.com; yangsongjie@std.uestc.edu.cn; zhangzp@uestc.edu.cn).
			
			Phee Lep Yeoh and Yonghui Li are with the School of Electrical and Information Engineering, University of Sydney, Sydney, NSW 2006, Australia (e-mail: phee.yeoh@sydney.edu.au; yonghui.li@sydney.edu.au).
	}}

	\maketitle
	
	\begin{abstract}
		
		This paper investigates the optimization of reconfigurable intelligent surface (RIS) in an integrated sensing and communication (ISAC) system. \red{To meet the demand of growing number of devices, power domain non-orthogonal multiple access (NOMA) is considered. However, traditional NOMA with a large number of devices is challenging due to large decoding delay and propagation error introduced by successive interference cancellation (SIC). Thus, OMA is integrated into NOMA to support more devices}. We formulate a max-min problem to optimize the sensing beampattern \red{with constraints on communication rate}, through joint power allocation, active beamforming and RIS phase shift design. To solve the non-convex problem with a non-smooth objective function, we propose a low complexity alternating optimization (AO) algorithm, where a closed form expression for the intra-cluster power allocation (intra-CPA) is derived, and penalty and successive convex approximation (SCA) methods are used to optimize the beamforming and phase shift design. Simulation results show the effectiveness of the proposed algorithm in terms of improving minimum beampattern gain (MBPG) compared with other baselines. Furthermore, the trade-off between sensing and communication is analyzed and demonstrated in the simulation results.
		
	\end{abstract}
	
	% Note that keywords are not normally used for peer review papers.
	
	\IEEEpeerreviewmaketitle
	\section{Introduction}
	\vspace{-0.1cm}
	\IEEEPARstart{I}{ntegrated} sensing and communication (ISAC) has attracted much attention in future wireless networks due to its wide application scenarios such as autonomous driving, indoor localization, and extended reality (XR). Sharing the same frequency band and hardware platform, ISAC is designed to perform joint sensing and communication to enhance the spectral efficiency (SE) and reduce hardware costs \cite{Survey}. One of the key challenges of ISAC is the waveform design. \red{In \cite{CoeX2Joint}, the authors investigated both separated and shared deployment for radar and communication antennas.} \red{Authors in \cite{ICASSP} maximized the minimum received signal-to-interference-plus-noise-ratio (SINR) in a joint radar and communication system with constraint on the radar sensing performance.}
	
	A promising technology to improve transmit waveforms in sixth generation mobile networks (6G) is the reconfigurable intelligent surface (RIS). \red{A large number of independently controllable reflecting elements on RIS} can provide extra degrees of freedom (DoFs) to reconfigure the wireless environment \cite{AerialRIS,JZJ,WJ_TWC}. Since electromagnetic wave propagation usually experiences large path loss and non-line-of-sight (NLoS) links, RIS has been exploited in ISAC to improve both radar and communication functionalities. In \cite{JSTSP}, RIS was employed in a dual functional radar-communication (DFRC) system to maximize the radar output SINR \red{while guaranteeing the quality of service (QoS) of the communication users.} Authors in \cite{JSAC_RIS} studied a double-RIS assisted communication radar coexistence system under high and low radar transmit power scenarios respectively. 
	
	%	Due to the explosive growth of in the number of accesses, 6G has higher demands on the SE. As a key technology to boost the SE \cite{Joint_TxRx}, non-orthogonal multiple access (NOMA) is employed in ISAC, enabling multiple users to share the same time and frequency resources by superposition coding and successive interference cancellation (SIC). In \cite{MultiUni}, radar was integrated into NOMA aided multicast-unicast communication by transmitting the superimposed signal to serve the radar- and communication- users and provide radar sensing simultaneously. In \cite{ICC_IC}, interference cancellation was investigated in NOMA empowered DFRC, where superimposed communication and sensing signals were jointly designed to match the ideal beampattern.
	Due to the explosive growth in the number of devices in ISAC systems such as automatic driving and internet of things (IoT) networks, higher demands on the SE becomes a key issue. Moreover, severe interference is another challenge in future ISAC. Although this can be alleviated by multi-antenna technologies, limited spatial degrees of freedom (DoFs) leads to insufficient support \red{for sensing targets and serving users} \cite{10122520}. As a key technology to boost the SE and mitigate the influence of limited spatial DoFs \cite{Joint_TxRx}, non-orthogonal multiple access (NOMA) is employed in ISAC, enabling multiple users to share the same time and frequency resources by superposition coding and successive interference cancellation (SIC) \cite{MultiUni}. \red{In \cite{9668964}, weighted sum throughput and effective sensing power was maximized by beamforming optimization. In \cite{9927490}, the authors explored the secrecy rate maximization in NOMA aided ISAC systems.} However, multiple users operating in pure NOMA mode can result in large sequential decoding delay, severe propagation error and high bandwidth requirements \cite{IOTJ_HMA}. To this end, orthogonal multiple access (OMA) and NOMA schemes can be combined by dividing the available radio resources (i.e. time or frequency) into orthogonal sub-resource blocks (RBs). In hybrid OMA-NOMA networks, users are divided into clusters, where each cluster is assigned with one sub-RB, serving the users in one cluster by power domain multiplexing \cite{MultiCarrierNOMA,TDMA_NOMA}.
	
	In this paper, a RIS aided hybrid NOMA-ISAC network is studied by maximizing the minimum beampattern gain (MBPG) at the sensing target angles, under the constraints of the communication users' rate threshold, transmit power budget and reflecting phase shifts. However, the formulated problem is non-convex and difficult to solve. To decouple the variables, alternating optimization (AO) algorithm is used to decompose the problem into three subproblems. For intra-cluster power allocation (intra-CPA), we provide a closed form expression for the optimal solution. For the active and passive beamforming problems, we tackle the non-convex constraints with successive convex approximation (SCA) method. Also, penalty method is utilized to deal with the unit modulus equality constraint for passive beamforming optimization. Finally, numerical simulations are performed to evaluate the performance of the proposed algorithm, which outperforms the other three baselines significantly. The trade-off between sensing and communication performance is also shown in the numerical results.

	\section{System Model and Problem Formulation}
	
	\subsection{Communication and Radar System Model}
	
%	\begin{figure}[!t]
%		\centering
%		\includegraphics[width=0.5\linewidth]{SysModel.png}
%		\caption{System model of the RIS assisted NOMA empowered ISAC system}
%		\label{model}
%		\vspace*{-0.5\baselineskip}
%	\end{figure}
	
\begin{figure}[!t]
		\centering
		\includegraphics[width=0.6\linewidth]{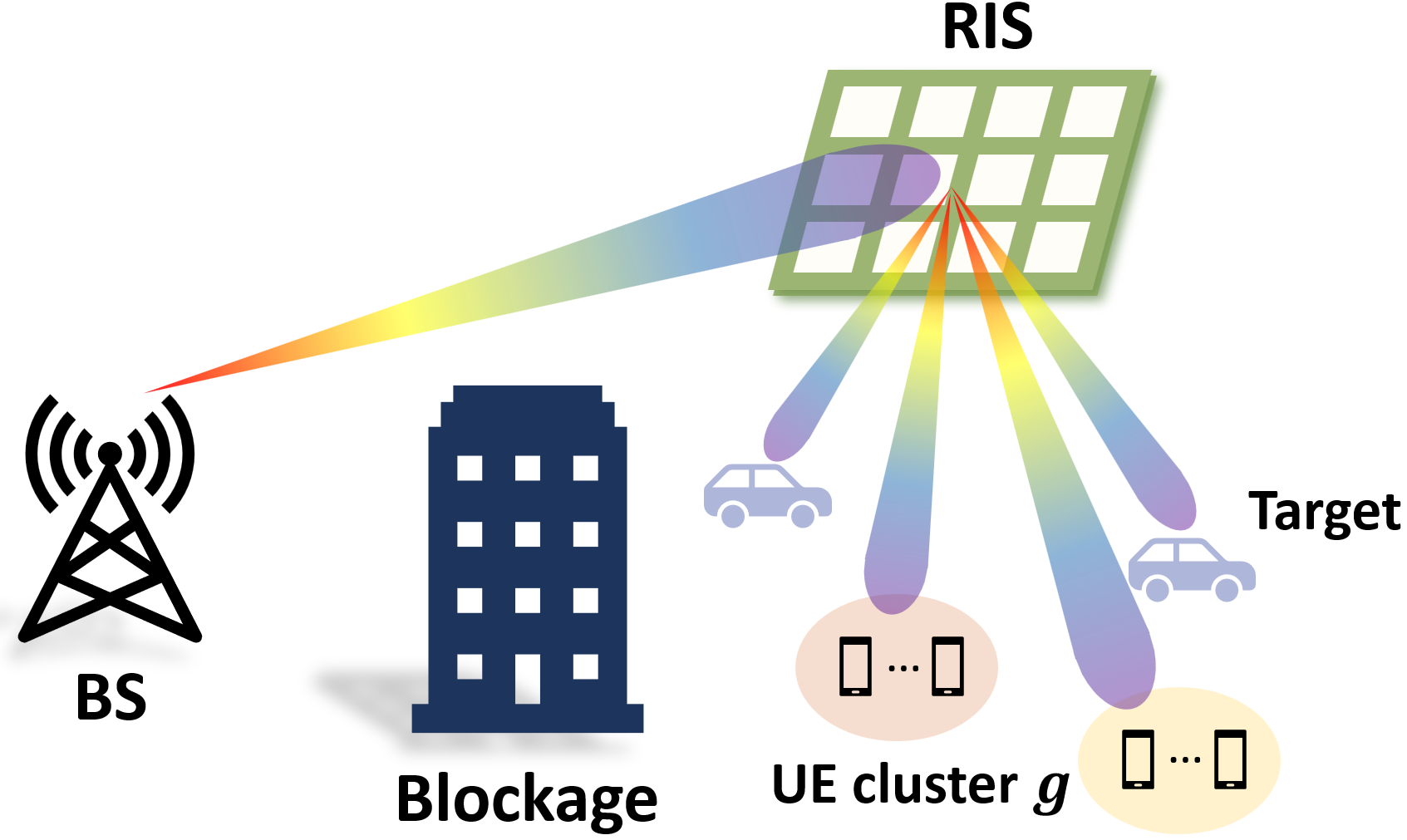}
		\caption{System model of the RIS assisted NOMA empowered ISAC system}
		\label{model}
%		\vspace*{-0.5\baselineskip}
	\end{figure}

	We consider a hybrid NOMA empowered ISAC system as shown in Fig. \ref{model}, where an $M$-antenna sensing and communication dual-functional base station (BS) serves $K$ single-antenna communication users (CommUs) and detect $L$ radar targets  simultaneously, aided by a RIS configured with $N_s$ passive reflecting elements. \red{The direct links from the BS to the CommUs and the targets are assumed to be blocked which motivates the use of RIS \cite{10122520,XY_TWC}}. The CommUs are divided into $G$ clusters, each allocated with one sub-RB. The radio resource allocation schemes can be found in \cite{IOTJ_HMA,MultiCarrierNOMA,TDMA_NOMA}, which is beyond the scope of this paper. Cluster $g$ contains $P_g$ CommUs, satisfying $\sum_{g=1}^G P_g = K$. We use $\mathcal{U}(g,p)$ to represent the $p$-th CommU in the $g$-th cluster. The transmitted signal of the CommU $\mathcal{U}(g,p)$ is denoted as $s_{g,p}$ satisfying $\mathbb{E}(|s_{g,p}|^2) = 1$. Assume that the CommUs and radar targets are distributed in similar directions of the RIS coverage range, so the communication waveform can be used for both communication and sensing \cite{CoeX2Joint}. Thus, the transmitted signal for the $g$-the cluster can be expressed as
	% \begin{small}
	\begin{equation}
		\mathbf x_g  =\mathbf w_g\sum_{p=1}^{P_g} \sqrt{\alpha_{g,p}}s_{g,p},
	\end{equation}
	% \end{small}%
	where $\mathbf w_g \in \mathbb C^{M\times 1}$ denotes the beamforming vector for cluster $g$, and $\alpha_{g,p}$ denotes the intra-cluster power allocated to $\mathcal U(g,p)$. Let $\mathbf H\in\mathbb C^{N_s\times M}$ and $\mathbf h_{g,p}^H \in \mathbb C^{1\times N_s}$ represent the wireless channels from the BS to the RIS, and from the RIS to $\mathcal U(g,p)$, respectively. The channels state information (CSI) is assumed to be perfectly known, where RIS-aided channel estimation methods can be found in \cite{YSJTVT}. Since the sub-RBs for different clusters are orthogonal, the received signal at $\mathcal U(g,p)$ can be expressed as 
	% \begin{small}
	\begin{equation}
		\begin{split}
			y_{g,p} =  &\underbrace{\tilde{\mathbf h}_{g,p}^H\mathbf w_g\sqrt{\alpha_{g,p}}s_{g,p}}_\text{desired signal}~  +~\underbrace{n_{g,p}}_\text{noise} \\
			+ &\underbrace{\tilde{\mathbf h}_{g,p}^H\mathbf w_g \left( \sum_{j=1}^{p-1}\sqrt{\alpha_{g,j}}s_{g,j} + \sum_{j=p+1}^{P}\sqrt{\alpha_{g,j}}s_{g,j} \right)}_\text{intra-cluster interference},
			\label{received_sig}
		\end{split}
	\end{equation}
	% \end{small}%
	where $\tilde{\mathbf h}_{g,p}^H = \mathbf h_{g,p}^H\mathbf\Phi\mathbf H$ is the cascaded channel between the BS and CommU $\mathcal U(g,p)$ dependent on the RIS reflecting matrix $\mathbf\Phi = \diag\{ e^{j\phi_1}, \cdots, e^{j\phi_{N_s}}\} \in \mathbb C^{N_s\times N_s}$. The noise term represents the additive white Gaussian noise (AWGN) $n_{g,p} \sim \mathcal{CN}(0,\sigma_0^2)$.

	Applying successive interference cancellation (SIC) in downlink NOMA \cite{dll_noma}, the decoding order in cluster $p$ is determined by the effective channel gains. Assume that $\vert \tilde{\mathbf h}_{g,1}^H\mathbf w_g \vert \le \vert \tilde{\mathbf h}_{g,2}^H\mathbf w_g \vert \le \cdots \le \vert \tilde{\mathbf h}_{g,Pg}^H\mathbf w_g\vert$, and the signal of $\mathcal U(p,j)$ can be decoded by $\mathcal U(p,g)$ for $j \le g$, and then removed from the $y_{g,p}$. Thus, the received SINR of $\mathcal U(p,g)$ can be written as
	% \begin{small}
	\begin{equation}
		\gamma_{g,p} = \frac{\vert \tilde{\mathbf h}^H_{g,p}\mathbf w_g \vert^2 \alpha_{g,p}}{ \vert \tilde{\mathbf h}^H_{g,p}\mathbf w_g \vert^2 \sum_{j=p+1}^{P_g}\alpha_{g,j} + \sigma_0^2},
	\end{equation}
	% \end{small}%
	and the corresponding data rate is $R_{g,p} = \log_2(1+\gamma_{g,p})$.
	
	Since the communication signal is a dual-functional waveform, the transmit beampattern for radar sensing can be expressed as
%	\begin{small}
	\begin{equation}
		P(\theta, \mathbf w_g, \mathbf \Phi) = \mathbf a^H(\theta)\mathbf\Phi\mathbf H\left(\sum_{g=1}^G \mathbf R_g\right)\mathbf H^H\mathbf\Phi^H\mathbf a(\theta),
	\end{equation}
%	\end{small}%
	where $\mathbf R_g = \mathbb E(\mathbf x\mathbf x^H) = \mathbf w_g\mathbf w_g^H$ is the covariance matrix of the transmitted signal. $\theta$ is the target detection angle over the coverage range of $[-\pi/2, \pi/2]$, and $\mathbf a(\theta) = \left[ 1, e^{j\frac{2\pi d}{\lambda}\sin\theta}, \cdots, e^{j\frac{2\pi (N_s-1)d}{\lambda}\sin\theta} \right]^T \in \mathbb C^{Ns\times 1}$ is the steering vector of the RIS, with $\lambda$ denoting the carrier wavelength, and $d = \lambda/2$ denotes the space between the adjacent elements.
	\subsection{Problem Formulation}
	
	In this paper, we aim to optimize the sensing performance with constraints on QoS through jointly optimizing the power allocation, active beamforming and RIS phase shift design. The objective function to measure the sensing performance is to maximize the MBPG at the target angles. Thus, the optimization problem can be formulated as
%	\begin{small}
		\begin{align}
			\max_{\alpha_{g,p},\atop \mathbf w_g, \mathbf\Phi} & \min_{l\in\mathcal L} P(\theta_l,\mathbf w_g,\mathbf\Phi) \label{P1obj} \\
			\text{s. t. }
			& R_{g,p} \ge r^{th}_{g,p},\ \forall g, p \label{Cons_SINR}\tag{\ref{P1obj}a}\\
			& \sum_{g=1}^G\Vert \mathbf w_g\Vert_2^2 \le P_t, \label{Cons_Txpow}\tag{\ref{P1obj}b}\\
			& \sum_{p=1}^{P_g} \alpha_{g,p} = 1,\  \forall g, \label{Cons_PowAlloc}\tag{\ref{P1obj}c}\\
			& \alpha_{g,p} \in (0, 1),\ \forall g, p \label{Cons_PowAlloc01}\tag{\ref{P1obj}d}\\
			& \phi_{n_s} \in [0, 2\pi),\ \forall n_s,  \label{Cons_phase} \tag{\ref{P1obj}e}
		\end{align}
%	\end{small}%
	where $\mathcal L$ is the set of the desired angles of the targets. (\ref{Cons_SINR}) is to guarantee the minimum data rate for CommUs. (\ref{Cons_Txpow}) sets the transmit power budget, while (\ref{Cons_PowAlloc}) and (\ref{Cons_PowAlloc01}) are the constraints for intra-cluster power allocation (intra-CPA). (\ref{Cons_phase}) is the constraint for RIS phase shifts. However, this problem is difficult to solve because it is non-convex with highly coupled variables. Thus, we propose an AO based iterative algorithm to solve it in Section \ref{III}.
	
	\section{Joint Optimization Algorithm}\label{III}
	
	In this section, to decouple the variables, we decompose the non-convex problem (\ref{P1obj}) into three sub-problems, and update the corresponding variable with others fixed in each iteration.
	
	\subsection{ Intra-cluster Power Allocation Optimization }
	
	In this sub-problem, we first optimize the intra-CPA $\alpha_{p,g}$, provided that the active and passive beamforming design $\mathbf w_g$ and $\mathbf\Phi$ are given. Since $\alpha_{g,P_g}$ is only related to the communication part, the problem (\ref{P1obj}) becomes a feasibility-check problem. To this end, we consider to maximize the communications performance evaluated by the achievable sum rate, and the problem can be reformulated as
%	\begin{small} 
	\begin{align}
		\max_{\alpha_{g,p}} & \sum_{g=1}^G\sum_{p=1}^{P_g} R_{g,p}  \label{P2obj} \\
		\text{s. t. } & (\ref{Cons_SINR}), (\ref{Cons_PowAlloc}), (\ref{Cons_PowAlloc01}). \nonumber
	\end{align} 
%	\end{small}%
	To maximize the sum rate of all $K$ CommUs, the intra-CPA problem becomes maximizing the sum rate in each cluster, respectively.
	
	\emph{\textbf{Lemma 1}: For all $ p_0 \in \{1,...,P_g-1\}$, if all $\alpha_{g,p}, p \in \{1,...,p_0-1,p_0+2,...,P_g\}$ are fixed, $R_g$ is non-increasing with $\alpha_{g,p_0}$ increasing.  }
	
	\emph{\textbf{Proof:}} please see \textbf{Appendix}.

	%%% 要改
	According to \emph{\textbf{Lemma 1}}, to maximize the sum rate in each cluster, the minimum amount of power should be allocated to the weak CommUs, namely $\mathcal U(g,1),...,\mathcal U(g,P_g-1)$ to satisfy their minimum rate constraints, while the maximum possible power should be allocated to the strong CommU $\mathcal U(g,P_g)$. Hence, in any cluster $g$, the intra-CPA policy should find the solution to the following equations,
%	\begin{small} 
	\begin{equation}
		\left\{
		\begin{aligned}   
			& R_{g,p} = r_{g,p}^{th},\ 1\le p \le P_g-1, \\
			& \sum_{g=1}^{P_g}\alpha_{p,g} = 1.
		\end{aligned}\right.
	\end{equation}
%	\end{small} %
	This is equivalent to solving the following equations,
%	\begin{small}
		\begin{subequations}
			\begin{align}
				&\log_2\left(1+ \frac{\vert \tilde{\mathbf h}_{g,1}^H\mathbf w_g \vert_2^2 \alpha_{g,1}}{\vert \tilde{\mathbf h}_{g,1}^H\mathbf w_g \vert_2^2\sum_{i=2}^{P_g}\alpha_{g,j} + \sigma_0^2} \right)= r_{g,1}^{th}, \\ 
				& \qquad\qquad\qquad\vdots \\
				&\log_2\left(1+ \frac{\vert \tilde{\mathbf h}_{g,P_g-1}^H\mathbf w_g \vert_2^2\alpha_{g,P_g-1}}{\vert \tilde{\mathbf h}_{g,P_g-1}^H\mathbf w_g \vert_2^2\alpha_{g,Pg} + \sigma_0^2} \right)= r_{g,P_g-1}^{th}, \\
				&\qquad \qquad\qquad\sum_{p=1}^{P_g}\alpha_{g,p} = 1.
			\end{align}
		\end{subequations}
%	\end{small}%
	Accordingly, the closed form expression for the optimal power allocation in iteration $i$ can be obtained as
	\begin{small}
		\begin{subequations}
			\begin{align}
				&\alpha_{g,1}^{(i)} = \frac{\gamma_{g,1}^{th}}{1 + \gamma_{g,1}^{th}}\left( 1 + \frac{\sigma_0^2}{\vert \tilde{\mathbf h}_{g,1}^H\mathbf w_g \vert_2^2} \right), \\	
				& \qquad\quad\qquad\vdots \\
				&\alpha_{g,P_g-1}^{(i)} = \frac{\gamma_{g,P_g-1}^{th}}{1 + \gamma_{g,P_g-1}^{th}}\left( 1 -\sum_{j=1}^{P_g-2}\alpha_{g,j} + \frac{\sigma_0^2}{\vert \tilde{\mathbf h}_{g,P_g-1}^H\mathbf w_g \vert_2^2} \right), \\
				&\alpha_{g,P_g}^{(i)} = 1 - \sum_{p=1}^{P_g-1} \alpha_{g,p},
			\end{align} \label{Intra-CPA}
		\end{subequations}
	\end{small}%
	where $\gamma_{g,p} = 2^{r_{g,p}^{th}} -1 $ denotes the SINR threshold for $\mathcal U(g,p)$.
	
	\setParDis
	\subsection{Active Beamforming Optimization}
%	\vspace{-0.1cm}
	With fixed $\alpha_{g,p}^{(i)}$ and $\mathbf\Phi^{(i)}$, the optimization problem can be further reformulated as
%	\begin{small}
	\begin{align}
		\max_{\mathbf w_g} & \min_{l\in\mathcal L} P(\theta_l,\mathbf R_g,\mathbf\Phi) \label{P3obj} \\
		\text{s. t. }
		& (\ref{Cons_SINR}),(\ref{Cons_Txpow}) \notag,
	\end{align}
%	\end{small}%
	which is a non-convex problem with non-smooth objective function and non-convex constraints. To deal with the max-min objective function, we first introduce an auxiliary variable $t$, and the problem can be re-expressed as
%	\begin{small}
	\begin{align}
		\max_{\mathbf w_g,t}\; & t\label{P4obj} \\
		\text{s. t. } & P(\theta_l,\mathbf w_g,\mathbf\Phi) \ge t, \forall l\in\mathcal L, \label{Cons_BP} \tag{\ref{P4obj}{a}} \\
		& (\ref{Cons_SINR}),(\ref{Cons_Txpow}) \notag.
	\end{align}
%	\end{small}%
	% 	Generally, this problem can be solved by semidefinite relaxation (SDR) method by directly optimizing the rank-one covariance matrix of the transmitted signal $\mathbf R_g$ \cite{9124713}. 
	Since $P(\theta_l,\mathbf w_g,\mathbf\Phi)$ can be lower bounded by its first order Taylor expansion, we use the successive convex approximation (SCA) method and the non-convex constraint (\ref{Cons_BP}) can be rewritten as
%	\begin{small}
	\begin{equation}
		\sum_{g=1}^G \left(2\mathcal Re\left\{\mathbf w_g^{(i)H}\mathbf b_l\mathbf b_l^H\mathbf w_g \right\} - \left|\mathbf b_l^H\mathbf w_g^{(i)}\right|^2 \right) \ge t, \label{Cons_cvx_BP}
	\end{equation}
%\end{small}%
	where $\mathbf b_l^H = \mathbf a^H(\theta_l)\mathbf\Phi\mathbf H$, and $(i)$ denotes the value obtained in the last iteration $i$.
	
	For constraint (\ref{Cons_SINR}), we first rewrite it as 
%	\begin{small}
	\begin{equation}
		\left|\tilde{\mathbf h}_{g,p}^H\mathbf w_g\right|^2 \ge \frac{\gamma_{g,p}^{th}}{\alpha_{g,p}} \left( \left\vert \tilde{\mathbf h}^H_{g,p}\mathbf w_g \right\vert_2^2 \sum_{j=p+1}^{P_g}\alpha_{g,j} + \sigma_0^2 \right). \label{Cons_SINR_r1}
	\end{equation}
%\end{small}%
	Similarly, the left hand side of (\ref{Cons_SINR_r1}) can be approximated iteratively by
	\begin{equation}
		\begin{split}
			2\mathcal Re\left\{ \mathbf w_g^{(i)H}\tilde{\mathbf h}_{g,p}\tilde{\mathbf h}_{g,p}^H\mathbf w_g  \right\} \red{- 	\left|\tilde{\mathbf h}_{g,p}^H\mathbf w_g^{(i)}\right|^2} \\ \ge \frac{\gamma_{g,p}^{th}}{\alpha_{g,p}}  \left( \left\vert \tilde{\mathbf h}^H_{g,p}\mathbf w_g \right\vert_2^2 \sum_{j=p+1}^{P_g}\alpha_{g,j} + \sigma_0^2 \right). \label{Cons_cvx_SINR}
		\end{split}
	\end{equation}%
	
	Hence, the active beamforming optimization problem can be rewritten as 
%	\begin{small}
	\begin{align}
		\max_{\mathbf w_g, t}\; & t \label{P5obj} \\
		\text{s. t. } & (\ref{Cons_Txpow}), (\ref{Cons_cvx_BP}), (\ref{Cons_cvx_SINR}), \nonumber
	\end{align}
%\end{small}%
	which is a convex problem that can be solved efficiently by standard convex solvers such as CVX \cite{boyd2004convex}.\vspace{-0.6cm}
	
	\subsection{RIS Phase Shift Optimization}
%	\vspace{-0.3cm}
	With fixed $\alpha_{g,p}$ and $\mathbf w_g$, we can transform the problem into a more tractable form by rewriting the cascaded RIS channel into $\tilde{\mathbf h}_{g,p}^H = \mathbf h_{g,p}^H\mathbf\Phi\mathbf H = \mathbf v^H\diag\{\mathbf h_{g,p}\}\mathbf H$, where $\mathbf v^H = \left[e^{j\phi_1}, \cdots, e^{j\phi_{N_s}}\right]$ denotes the RIS phase shift vector. Accordingly, the phase constraint (\ref{Cons_phase}) becomes a unit modulus constraint as follows,
%	\begin{small}
	\begin{equation}
		\vert v_{n_s}\vert = 1,\ n_s \in\{1,...,N_s\}. 
		\label{Cons_unit_modulus}
	\end{equation}
%\end{small}%
	Similarly, the radar sensing beampattern can be rearranged as $P(\theta_l,\mathbf w_g,\mathbf v) = \mathbf v^H\mathbf Q(\theta_l)\mathbf v$, where $\mathbf Q(\theta_l) = \diag\{\mathbf a(\theta_l)^H\}\mathbf H\left( \sum_{g=1}^G\mathbf w_g\mathbf w_g^H \right)\mathbf H^H\diag\{\mathbf a(\theta_l)\}$. Then, the sub-problem for optimizing $\mathbf v$ can be formulated as
%	\begin{small}
		\begin{align}
			\max_{\mathbf v, t}\; &t \label{P6obj} \\
			\text{s. t. } & \mathbf v^H\mathbf Q(\theta_l)\mathbf v \ge t, \label{Cons_v_BP} \tag{\ref{P6obj}a} \\
			& \log_2\left(1+\frac{\left\vert\mathbf v^H\diag\{\mathbf h_{g,p}\}\mathbf H\mathbf w_g\right\vert^2\alpha_{g,p}}{ \left\vert\mathbf v^H\diag\{\mathbf h_{g,p}\}\mathbf H\mathbf w_g\right\vert^2\sum_{j=p+1}^{P_g}\alpha_{g,p} }\right) \ge r_{g,p}^{th}, \label{Cons_v_SINR} \tag{\ref{P6obj}b} \\ 
			&(\ref{Cons_unit_modulus}), \nonumber 
		\end{align}
%	\end{small}%
	where the non-convex constraints (\ref{Cons_v_BP}) and (\ref{Cons_v_SINR}) can be similarly tackled by SCA algorithm as in the last section. In addition, we deal with the unit modulus constraint (\ref{Cons_unit_modulus}) by penalty-based method. Specifically, we reformulate the problem to be a penalized version as 
	\begin{small}
		\begin{align}
			\max_{\mathbf v, t}\; & t + \mu\sum_{n_s=1}^{N_s} (|v_{n_s}|^2 -1)  \label{P6obj} \\ 
			\text{s. t. } & 2\mathcal Re\{ \mathbf v^{(i-1)H}\mathbf Q(\theta_l)\mathbf v \}  - \mathbf v^{(i-1)H}\mathbf Q(\theta_l)\mathbf v^{(i-1)} \ge t, \label{Cons_cvx_vBP} \tag{\ref{P6obj}{a}} \\
			& z_g(\mathbf v) \ge \frac{\gamma_{g,p}^{th}}{\alpha_{g,p}}\left(  \left\vert\mathbf v^H\diag\{\mathbf h_{g,p}\}\mathbf H\mathbf w_g\right\vert^2\sum_{j=p+1}^{P_g}\alpha_{g,p}  \right), \label{Cons_cvx_vSINR} \tag{\ref{P6obj}{b}} \\ 
			& |v_{n_s}| <= 1, \label{Cons_v}\tag{\ref{P6obj}{c}}
		\end{align}
	\end{small}%
	where $\mu$ is a large positive constant, and $z_g(\mathbf v) = 2\mathcal Re\{ \mathbf v^{(i-1)H}\diag\{\mathbf h_{g,P_g}^H\}\mathbf H\mathbf w_g\mathbf w_g^H\mathbf H^H\diag\{\mathbf h_{g,P_g}\}\mathbf v \}  - \mathbf v^{(i-1)H}\diag\{\mathbf h_{g,P_g}^H\}\mathbf H\mathbf w_g\mathbf w_g^H\mathbf H^H\diag\{\mathbf h_{g,P_g}\}\mathbf v^{(i-1)}$ is the lower bound of $\vert \mathbf v^H\diag\{\mathbf h_{g,P_g}^H\}\mathbf H \mathbf w_g\vert^2 $. Furthermore, we iteratively approximate the non-convex objective function by its first order Taylor expansion, and thus the problem (\ref{P6obj}) can be transformed as
%	\begin{small}
	\begin{align}
		\max_{\mathbf v, t}\; & t + 2\mu\sum_{n_s=1}^{N_s} \mathcal Re\{v_{n_s}^{(i-1)*}(v_{n_s} - v_{n_s}^{(i-1)})\}  \label{fi_obj} \\ 
		\text{s. t. } & (\ref{Cons_cvx_vBP}), (\ref{Cons_cvx_vSINR}), (\ref{Cons_v}),  \nonumber
	\end{align}
%	\end{small}%
	which is a convex problem. Note that two-layer iteration is adopted in this problem, where in the outer layer, the penalty factor $\mu$ is first set as a small value to find a good starting point, then updated until sufficiently large. In the inner layer, $\mu$ is fixed, and (\ref{fi_obj}) is solved iteratively to update $\mathbf v$ and $t$. 
	\begin{algorithm}[t]  
		\caption{Joint Power Allocation, Beamforming and Phase Shift Optimization for the RIS aided hybrid NOMA ISAC}  
		\begin{algorithmic}[1]  
			\State \textbf{Initialize} $\mathbf w_g^{(0)}, \forall g, \mathbf v^{(0)} $ and $\mu^{(0)}$. Set iteration index $i = 0$.
			\Repeat
			\State \textbf{Update} intra-CPA coefficients $\alpha_{g,p}$ by (\ref{Intra-CPA}).
			\Repeat 
			\State \textbf{Update} active beamforming vector $\mathbf w_g$ and auxiliary variable $t$ by solving problem (\ref{P5obj}).
			\Until{Convergence.}
			\Repeat$\,$ outer loop:
			\State \textbf{Enlarge} penalty factor $\mu$.
			\Repeat$\,$ inner loop:
			\State \textbf{Update} RIS phase shift vector $\mathbf v$ and auxiliary variable $t$ by solving problem (\ref{fi_obj}).
			\Until{Convergence.}
			\Until{$\mu$ is sufficiently large.}
			\State Set $i = i + 1$.
			\Until{Convergence.}
		\end{algorithmic}  
	\end{algorithm}

\section{\red{Complexity Analysis}}

\red{Overall, the proposed algorithm for solving the problem is summarized in \textbf{Algorithm 1}. Since the closed-form solution to the intra-CPA problem is given, the main computational complexity of Algorithm 1 is due to the active beamforming and RIS phase shifts optimization based on SCA method. In problem (\ref{P5obj}), there are $2GM+1$ real variables, then the computational complexity solving the active beamforming problem is $\mathcal O(I_1(2GM+1)^{3.5}\log_2(1/\epsilon))$, where $I_1$ is the number of iterations for this sub-problem, and $\epsilon$ is the accuracy of the SCA method. Similarly, the computational complexity for optimizing the RIS phase shifts can be expressed as $\mathcal O(I_mI_2(2N_s+1)^{3.5}\log_2(1/\epsilon))$, where $I_m$ and $I_2$ denote the number of iterations for the outer and inner loop, respectively. Therefore, the total computation complexity for the proposed joint optimization algorithm is $\mathcal O(T(I_1(2GM+1)^{3.5}\log_2(1/\epsilon)+I_mI_2(2N_s+1)^{3.5}\log_2(1/\epsilon)))$.

}
 
	% \vspace{-0.3cm}
	\section{Numerical Results}
%	\vspace{-0.3cm}
		\begin{figure*}[t]
		\centering
		\subfigure[Minimum beampattern gain at interested angles versus the number of reflecting elements.]{
			\begin{minipage}[t]{0.32\linewidth}
				\centering
				\includegraphics[width=1\linewidth]{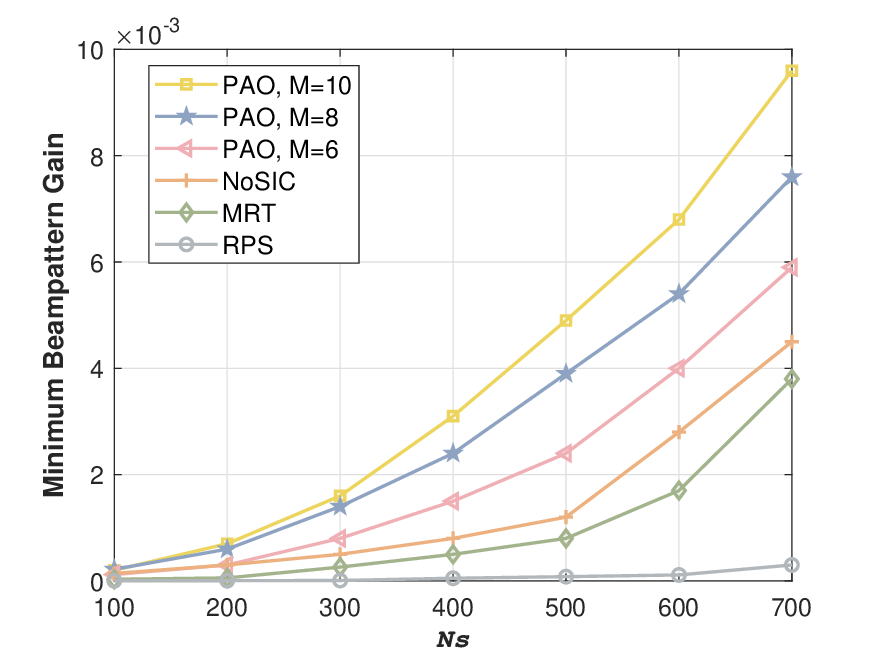}
				\label{BPvsNs}
			\end{minipage}
		}
		\subfigure[Minimum beampattern gain at interested angles versus the transmit power budget.]{
			\begin{minipage}[t]{0.32\linewidth}
				\centering
				\includegraphics[width=1\linewidth]{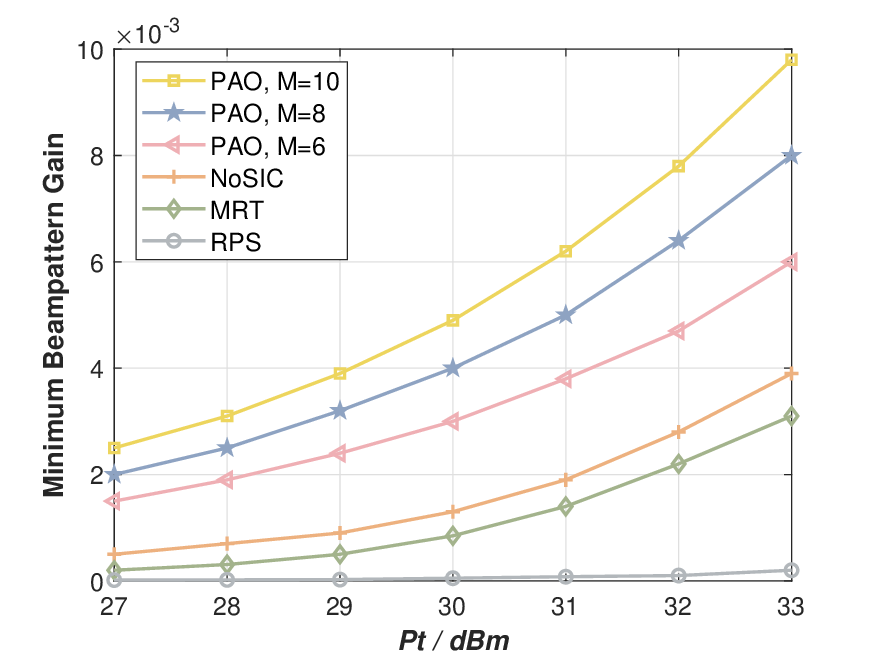}
				\label{BPvsPt}
			\end{minipage}
		}
		\subfigure[Beampattern gain along the sensing angles.]{
			\begin{minipage}[t]{0.3\linewidth}
				\centering
				\includegraphics[width=1\linewidth]{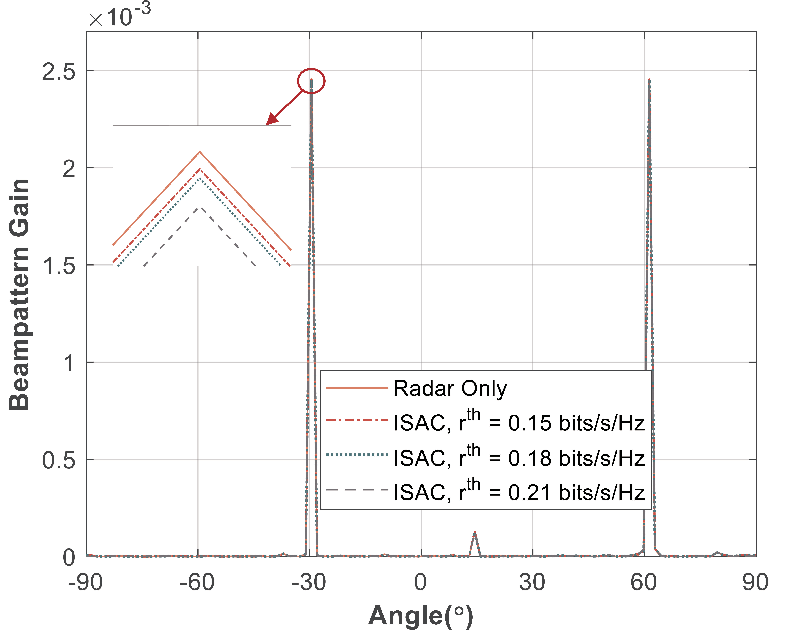}
				\label{BPgain}
			\end{minipage}
		}
		\caption{Simulation results}
		\label{Res}
	\end{figure*}
	In this section, the numerical simulation results are provided in Fig. \ref{Res} to evaluate the performance of the proposed AO based algorithm to solve the RIS aided hybrid NOMA-ISAC system. We consider a two-dimensional plane to describe the deployment of the system. The dual functional BS located at $(-20, 0)$m is configured with $M = 8$ transmit antennas unless otherwise stated. The RIS is at the origin of the coordinate system. We assume that there are $G = 2$ CommU clusters, with $P_1 = 3$ and $P_2 = 2$ CommUs. The CommUs in cluster $1$ are located at $(-20, 25)\text{m},~(-10, 9)\text{m},~(-5,5)\text{m}$ and the CommUs in cluster $2$ are located at $(14,18)\text{m},~(8,8)\text{m}$, respectively. Suppose there are two sensing target angles from the RIS at $-30^\circ$ and $60^\circ$, respectively. 
	
	The communication channels from the BS to the RIS, and from the RIS to the CommUs are assumed to follow the Rician fading distribution, with a Rician factor of $3$ dB. The path loss is computed as $L = C_0d^{-a}$, where $C_0 = -30$ dB is the reference path loss at distance $d_0 = 1$ m, $d$ denotes the corresponding path distance, and $a$ denotes the path loss exponent. The noise power is set as $\sigma_0^2 = -90$ dBm. The sensing range is $[-90^\circ, 90^\circ]$ at the RIS. 
	
	The impact of the number of RIS elements on sensing performance is evaluated in Fig. \ref{BPvsNs}. The BS transmit power is set as $P_t = 30$ dBm, the minimum rate for each CommU is set as $r_{g,p}^{th} = 0.144$ bits/s/Hz. It can be seen that the MBPG at the interested sensing angles increases quickly with the growing number of RIS elements $N_s$ from $100$ to $700$. This is because more RIS elements provide more reflecting paths, and enable higher passive beamforming gain. We compare the performance of the proposed AO based (PAO) algorithm with other baselines: i) hybrid NOMA without the assistance of SIC (NoSIC) ii) maximum ratio transmission (MRT) for active beamforming design and iii) random RIS phase shift scheme (RPS). The figure shows that PAO achieves significantly higher beampattern gain at the target angles compared with the other three baselines. Moreover, we analyze the impact of the number of transmit antennas $M$ on the sensing performance, and find that more transmit antennas can improve the sensing performance, since higher beamforming gain can be achieved at both the CommUs and sensing targets.
	
	The MBPG versus the transmit power $P_t$ is illustrated in Fig. \ref{BPvsPt}. Obviously, the MBPG increases drastically with larger $P_t$ from $27$ to $33$ dBm, since higher received signal power can be reached to satisfy the rate constraint for CommUs, and thus more extra power is used for sensing. Also, the PAO algorithm outperforms the other three baselines, and more transmit antennas provide better performance in terms of the MBPG, which is consistent with the comparisons in Fig. \ref{BPvsNs}.
	
	The BPG versus different angles within the sensing range is illustrated in Fig. \ref{BPgain}. Peaks appear at the target angles as expected. Moreover, the trade-off between communication and sensing can be seen in this figure, where we compare the sensing performance of the proposed ISAC system and radar only system. The peak of the BPG of radar only system outperforms ISAC system, and more strict communication requirement results in sensing performance loss. This is intuitive since CommUs share the signal power and beamforming gain to satisfy the quality of service demand.
	\vspace{-0.5cm}
	\section{Conclusion}
%	\vspace{-0.7cm}
	Overall, this paper analyzed a hybrid NOMA empowered ISAC with the assistance of RIS responsible for both sensing and communication. To focus on sensing performance, a problem maximizing MBPG at the interested angles were formulated with constraints on QoS for CommUs, transmit power budgets for active beamforming and phase adjusting range for RIS elements. An AO based low complexity algorithm was proposed to solve the non-convex problem. The optimal intra-CPA solution was derived as closed-form expression, the transmit beamforming was optimized iteratively with SCA algorithm, and the passive beamforming design was realized by penalty method with SCA. Simulation results highlighted the effectiveness and significant performance improvement of the proposed algorithm compared to benchmark schemes.
%	\vspace{-0.3cm}
	
    \appendices
    
 \section{Proof of Lemma 1}
	\begin{figure*}[b]
		\hrule
%		\begin{small}
			\begin{equation}
				\begin{split}
					& \frac{\partial R_g}{\partial \alpha_{g,p_0}} = \frac{\partial( R_{g,p_0} + R_{g,p_0+1} )}{\partial \alpha_{g,p_0}} \\
					&= \frac{1}{\ln 2}\frac{\vert \tilde{\mathbf h}^H_{g,p_0}\mathbf w_g \vert^2}{\vert \tilde{\mathbf h}^H_{g,p_0}\mathbf w_g \vert^2\left(\sum_{j=p_0+2}^{P_g}\alpha_{g,j} + \hat\alpha-\alpha_{g,p_0}  \right) + \sigma_0^2} - \frac{1}{\ln 2}\frac{\vert \tilde{\mathbf h}^H_{g,p_0+1}\mathbf w_g \vert^2}{\vert \tilde{\mathbf h}^H_{g,p_0+1}\mathbf w_g \vert^2\left(\sum_{j=p_0+2}^{P_g}\alpha_{g,j} + \hat\alpha-\alpha_{g,p_0}  \right) + \sigma_0^2} \\
					&=  \frac{1}{\ln 2} \frac{\sigma_0^2\left(\vert \tilde{\mathbf h}^H_{g,p_0}\mathbf w_g \vert^2 - \vert \tilde{\mathbf h}^H_{g,p_0+1}\mathbf w_g \vert^2  \right)}{\left(\vert \tilde{\mathbf h}^H_{g,p_0}\mathbf w_g \vert^2\left(\sum_{j=p_0+2}^{P_g}\alpha_{g,j} + \hat\alpha-\alpha_{g,p_0}  \right) + \sigma_0^2\right)\left(\vert \tilde{\mathbf h}^H_{g,p_0+1}\mathbf w_g \vert^2\left(\sum_{j=p_0+2}^{P_g}\alpha_{g,j} + \hat\alpha-\alpha_{g,p_0}  \right) + \sigma_0^2\right)}
				\end{split}\label{Deriv}\tag{24}
			\end{equation}
%		\end{small}%
	\end{figure*}
	Since $\alpha_{g,p}, p \in \{1,...,p_0-1,p_0+2,...,P_g\}$ are fixed, and due to constraint (\ref{Cons_PowAlloc}), $\alpha_{g,p_0+1}$ can be expressed as
%	\begin{small}
	\begin{equation}
		\begin{split}
			\alpha_{g,p_0+1} & = 1 - \sum_{j\neq p_0,p_0+1}\alpha_{g,j} - \alpha_{g,p_0} \\
			& = \hat{\alpha} - \alpha_{g,p_0},
		\end{split}
	\end{equation}
%\end{small}%
	where $\hat\alpha = 1 - \sum_{j\neq p_0,p_0+1}\alpha_{g,j} $ is a constant. Thus, the achievable sum rate of cluster $g$ can be rewritten as
%	\begin{small}
	\begin{equation}
		R_g = R_{g,p_0} + R_{g,p_0+1} + \text{Const},
	\end{equation}
%	\end{small}%
	where the sum rate of CommUs in cluster $g$ except from $\mathcal U(g,p_0)$ and $\mathcal U(g,p_0+1)$ is a constant. $R_{g,p_0}$ and $R_{g,p_0+1}$ can be rewritten as 
	\begin{small} % small不然要超出去
		\begin{equation}
			R_{g,p_0} = \log_2\left( 1 + \frac{\vert \tilde{\mathbf h}^H_{g,p_0}\mathbf w_g \vert^2 \alpha_{g,p_0}}{ \vert \tilde{\mathbf h}^H_{g,p_0}\mathbf w_g \vert^2 \left(\sum_{j=p+2}^{P_g}\alpha_{g,j} + \hat\alpha - \alpha_{g,p_0} \right)+ \sigma_0^2} \right),
		\end{equation}
	\end{small}%
	\begin{small}
	\begin{equation}
		R_{g,p_0+1} = \log_2\left( 1 + \frac{\vert \tilde{\mathbf h}^H_{g,p_0+1}\mathbf w_g \vert^2 (\hat\alpha - \alpha_{g,p_0})}{ \vert \tilde{\mathbf h}^H_{g,p_0+1}\mathbf w_g \vert^2 \sum_{j=p+2}^{P_g}\alpha_{g,j} + \sigma_0^2} \right).
	\end{equation}
	\end{small}%
	Hence, the first-order derivative of $R_g$ with respect to (w.r.t.) $\alpha_{g,p_0}$ can be computed as (\ref{Deriv}) at the bottom of this page.
	
	Recall the assumption for the decoding order that $\vert \tilde{\mathbf h}_{g,1}^H\mathbf w_g \vert \le \vert \tilde{\mathbf h}_{g,2}^H\mathbf w_g \vert \le \cdots \le \vert \tilde{\mathbf h}_{g,Pg}^H\mathbf w_g\vert$, $\frac{\partial R_g}{\partial \alpha_{g,p_0}} \le 0$, indicating that $R_g$ is non-increasing w.r.t. $\alpha_{g,p_0}$. Hence, \textbf{\emph{Lemma 1}} is proved.
	
    \section{\red{Convergence Analysis}}

    \red{To show the convergence of the overall algorithm, we show that the solution to each sub-problem converges to a stationary point.
    
    Since the active beamforming optimization and the phase shift optimization are both based on the SCA method, we evaluate the convergence behavior of active beamforming optimization, and that of RIS coefficient design can be derived in a similar manner.

	Note that (\ref{P3obj}) and (\ref{P4obj}) are equivalent, then we show (\ref{P4obj}) converges to a Karush-Kuhn-Tucker (KKT) point. Based on constraints (\ref{Cons_SINR_r1}) and (\ref{Cons_cvx_SINR}), we define the following function,
	\begin{equation}
		\mathcal F(\mathbf w_g) =  \frac{\gamma_{g,p}^{th}}{\alpha_{g,p}} \left( \left\vert \tilde{\mathbf h}^H_{g,p}\mathbf w_g \right\vert_2^2 \sum_{j=p+1}^{P_g}\alpha_{g,j} + \sigma_0^2 \right) - \left|\tilde{\mathbf h}_{g,p}^H\mathbf w_g\right|^2. \label{Fw} \tag{25}	\end{equation}  
	After the Taylor expansion, (\ref{Fw}) is upper bounded by
	\begin{align}
		&\hat{\mathcal F}(\mathbf w_g|\mathbf w_g^{(i)}) =  \frac{\gamma_{g,p}^{th}}{\alpha_{g,p}} \left( \left\vert \tilde{\mathbf h}^H_{g,p}\mathbf w_g \right\vert_2^2 \sum_{j=p+1}^{P_g}\alpha_{g,j} + \sigma_0^2 \right) \nonumber \\
		&\qquad \qquad \quad- 	2\mathcal Re\left\{ \mathbf w_g^{(i)H}\tilde{\mathbf h}_{g,p}\tilde{\mathbf h}_{g,p}^H\mathbf w_g  \right\} + \left|\tilde{\mathbf h}_{g,p}^H\mathbf w_g^{(i)}\right|^2.
		\label{Fw_hat} \tag{26}
	\end{align}
	The proposed algorithm can converge to a KKT point for the optimization in problem (\ref{P4obj}) if the following conditions are satisfied \cite{1978A}:
	\begin{gather}
		\mathcal F(\mathbf w_g) \le \hat{\mathcal F}(\mathbf w_g|\mathbf w_g^{(i)}), \tag{27} \label{Cond1} \\
		\frac{\partial \mathcal F(\mathbf w_g)}{\partial \mathbf w_g} = \frac{\partial \hat{\mathcal F}(\mathbf w_g|\mathbf w_g^{(i)})}{\partial \mathbf w_g}, \tag{28} \label{Cond2}
	\end{gather}
	where condition (\ref{Cond1}) is obviously, and condition (\ref{Cond2}) can be verified by deriving the partial derivatives of $\mathcal F(\mathbf w_g)$ and 
	$ \hat{\mathcal F}(\mathbf w_g|\mathbf w_g^{(i)})$. 
	
	Thus, the proposed algorithm also converges to a KKT point for problem (\ref{P3obj}).Furthermore, we have that the solution to the RIS phase shifts design converges. Because the proposed algorithms for each sub-problem converges to a local optimal, and the overall problem is bounded, the overall algorithm converges with sufficient number of iterations. 
    }
 
	\ifCLASSOPTIONcaptionsoff
	\newpage
	\fi

	\vspace{-0.5cm}
	\bibliographystyle{IEEEtran}
	\bibliography{ISAC_RIS}

\end{document}